\documentclass[aps,prl,twocolumn,groupedaddress,showpacs,showkeyst]{revtex4-1}

\usepackage{amsmath,amssymb}
\usepackage{graphicx}
\usepackage[usenames,dvipsnames]{xcolor}
\usepackage{enumitem}

\newcommand{\bt}[1]{{\mathbf #1}}

\begin{document}
\title{Robust energy transfer mechanism via precession resonance \\ in nonlinear turbulent wave systems}
\author{Miguel D. Bustamante}
\email[]{miguel.bustamante@ucd.ie}
\affiliation{Complex and Adaptive Systems Laboratory, School of Mathematical Sciences,
University College Dublin,
Belfield, Dublin 4, Ireland}
\author{Brenda Quinn}
\email[]{bquinn@post.tau.ac.il}
\affiliation{Complex and Adaptive Systems Laboratory, School of Mathematical Sciences,
University College Dublin,
Belfield, Dublin 4, Ireland}
\author{Dan Lucas}
\email[]{dan.lucas@ucd.ie}
\affiliation{Complex and Adaptive Systems Laboratory, School of Mathematical Sciences,
University College Dublin,
Belfield, Dublin 4, Ireland}

\date{\today}

\begin{abstract}
A robust energy transfer mechanism is found in nonlinear wave systems, which favours transfers towards modes interacting via triads with nonzero frequency mismatch, applicable in meteorology, nonlinear optics and plasma wave turbulence. We introduce the concepts of truly dynamical degrees of freedom and triad precession. Transfer efficiency is maximal when the triads' precession frequencies resonate with the system's nonlinear frequencies, leading to a collective state of synchronised triads with strong turbulent cascades at intermediate nonlinearity. Numerical simulations confirm analytical predictions.


\end{abstract}

\maketitle

\bibliographystyle{apsrev4-1}
\vspace{-2mm}
\noindent \textbf{Introduction.} A variety of physical systems of high technological importance consist of nonlinearly interacting oscillations or waves: nonlinear circuits in electrical power systems, high-intensity lasers, nonlinear photonics, gravity water waves in oceans, Rossby-Haurwitz planetary waves in the atmosphere, drift waves in fusion plasmas, etc \cite{rewienski2003trajectory,protopapas1997atomic,kibler2010peregrine,Hasselmann1962,hoskins1973stability,HasegawaMima1978}. These systems are characterised by extreme events that are localised in space and time and are
associated with strong nonlinear energy exchanges that dramatically alter the system's global behaviour.
One of the few consistent theories that deal with these nonlinear exchanges is classical wave turbulence theory~\cite{Hasselmann1962,Zakharov1992,Nazarenko2011}. This theory produces statistical predictions by making ad-hoc hypotheses on correlations of the evolving quantities and is valid in the limit of weak nonlinearity.  One example where this theory is widely used is in numerical prediction of ocean waves \cite{komen1996dynamics}.

This Letter addresses a new robust mechanism of strong energy transfers in real physical systems, precisely in the context where the hypotheses of classical wave turbulence theory do not hold, namely when the spatial domains have a finite size, when the amplitudes of the carrying fields are not infinitesimally small and when the linear wave timescales are comparable to the timescales of the nonlinear oscillations. The theory that deals with these energy exchanges is Discrete and Mesoscopic Wave Turbulence~\cite{Lvov2010,Korotkevich2005,Nazarenko2006,Kartashova2007b,Bustamante2009a,Bustamante2009b,Kartashova2011,Harris2012,Harris2013,Harper2012,Bustamante2013} and is still in development. Our results apply to a variety of systems, namely the nonlinear partial differential equations (PDEs) of classical turbulence, nonlinear optics, quantum fluids and magneto-hydrodynamics considered on bounded physical domains. For the sake of simplicity of presentation we discuss here the Charney-Hasegawa-Mima (CHM) equation \cite{charney1971geostrophic,HasegawaMima1978}, a PDE governing Rossby waves in the atmosphere and drift waves in inhomogeneous plasmas:
$$(\nabla^2 - F)\frac{\partial \psi}{\partial t}  + \beta \frac{\partial \psi}{\partial x} + \frac{\partial \psi}{\partial x} \frac{\partial \nabla^2\psi}{\partial y} - \frac{\partial \psi}{\partial y} \frac{\partial \nabla^2\psi}{\partial x} = 0,$$
where in the plasma case the wave field $\psi(\mathbf{x},t) (\in \mathbb{R})$ is the electrostatic potential, $F^{-1/2}$ is the ion Larmor radius at the electron temperature
and $\beta$ is a constant proportional to the mean plasma density gradient. We assume periodic boundary conditions: $\mathbf{x} \in [0,2\,\pi)^2.$ Decomposing the field in Fourier harmonics, $\psi(\mathbf{x},t) = \sum_{\mathbf{k} \in \mathbb{Z}^2} A_{\mathbf{k}}(t) \mathrm{e}^{\mathrm{i} \mathbf{k}.\mathbf{x}}$ with wavevector $\bt{k}=(k_x,k_y),$  the components $A_{\mathbf{k}}(t)\,,\quad \mathbf{k} \in \mathbb{Z}^2$ satisfy the evolution equation
\begin{eqnarray}
\label{db_dt}
\dot{A}_{\bt k} + i\,\omega_{\bt{k}}\,{A}_{\bt k} &=& \frac{1}{2}\sum\limits_{\textbf{k}_1, \textbf{k}_2 \in \mathbb{Z}^2} Z^{\bt k}_{\bt{k}_1 \bt{k}_2}\, \delta_{\textbf{k}_1 +\textbf{k}_2-{\bt k}}\, A_{\bt{k}_1}\, A_{\bt{k}_2},
\vspace{-1.5mm}
\end{eqnarray}
where $Z^{\bt k}_{\bt{k}_1 \bt{k}_2} = (k_{1x}k_{2y}-k_{1y}k_{2x})\frac{|\bt{k}_1|^2-|\bt{k}_2|^2}{|\bt{k}|^2+F}$ are the interaction coefficients, $\omega_{\bt{k}} = \frac{-\beta k_x}{k^2+F}$ are the linear frequencies and $\delta$ is the Kronecker symbol. Reality of $\psi$ implies  $A_{-\bt{k}} = A_{\bt{k}}^*$ (complex conjugate). Since the degree of nonlinearity in the PDE is quadratic, the modes $A_{\bt{k}}$ interact in triads. A triad is a group of any three spectral modes $A_{\bt{k}_1}(t), A_{\bt{k}_2}(t), A_{\bt{k}_3}(t)$ whose wavevectors satisfy $\mathbf{k}_1+ \mathbf{k}_2 = \mathbf{k}_3.$ The triad's \emph{linear frequency mismatch} is defined by $\omega^{{\bt k}_3}_{\bt{k}_1 \bt{k}_2} \equiv \omega_{\bt{k}_1} + \omega_{\bt{k}_2} - \omega_{\bt{k}_3}.$

Since any mode belongs to several triads, energy can be transferred nonlinearly throughout the intricate network or cluster of connected triads. In weakly-nonlinear perturbation theory, triad interactions with non-zero frequency mismatch can be eliminated via a near-identity transformation. However, at finite nonlinearity these interactions cannot be eliminated \emph{a priori} because they take part in the triad precession resonances presented below. As evidence for this Letter's timeliness, it was established recently that triads/quartets with nonzero frequency mismatch are responsible for most of the energy exchanges in real systems~\cite{Janssen2003,Smith2005,Annenkov2006}. Here we consider inertial-range dynamics, i.e.~no forcing and no dissipation are imposed on the system and enstrophy cascades to small scales respect enstrophy conservation.\\

\vspace{-2mm}

\noindent \textbf{Truly dynamical degrees of freedom.} An under-exploited formulation for a large class of PDEs with quadratic nonlinearity is to write evolution equations for the ``truly dynamical'' degrees of freedom \cite{Bustamante2009b,Lynch2004}. To this end we introduce the amplitude-phase representation: $A_{\bt{k}} = \sqrt{n_{\bt{k}}} \exp(i\,\phi_{\bt{k}}),$ where $n_{\bt{k}}$ is called the wave spectrum \cite{Nazarenko2011}. The spectrum is constrained by the exact conservation in time of $E = \sum\limits_{\textbf{k}\in \mathbb{Z}^2} (|\bt{k}|^2+F) n_{\bt{k}}$ (energy) and $\mathcal{E} = \sum\limits_{\textbf{k}\in \mathbb{Z}^2} |\bt{k}|^2 (|\bt{k}|^2+F) n_{\bt{k}}$ (enstrophy). In the context of CHM equation (Galerkin-truncated to $N$ wavevectors), the truly dynamical degrees of freedom are any $N-2$ linearly independent \emph{triad} phases $\varphi^{\bt{k}_3}_{\bt{k}_1 \bt{k}_2} \equiv \phi_{\bt{k}_1} + \phi_{\bt{k}_2} - \phi_{\bt{k}_3}$ \cite{Harper2012,da2013quadratic} and the $N$ wave spectrum variables $n_{{\bt k}}.$ These $2N-2$ degrees of freedom satisfy a closed system of evolution equations (the individual phases $\phi_{\bt{k}}$ are slave variables, obtained by quadrature):
\begin{eqnarray}
\label{eq:eom_n}
 \dot{n}_{\bt{k}} &=& {\displaystyle \sum_{\bt{k}_1,\bt{k}_2}} Z^{\bt{k}}_{\bt{k}_1 \bt{k}_2} \delta_{\bt{k} - \bt{k}_1 - \bt{k}_2} ({n_{\bt{k}}\,n_{\bt{k}_1}\,n_{\bt{k}_2}})^{\frac{1}{2}}\,\cos\varphi^{\bt{k}}_{\bt{k}_1 \bt{k}_2}\,,\\
\nonumber
 \dot{\varphi}^{\bt{k}_3}_{\bt{k}_1 \bt{k}_2} &=& \sin{\varphi}^{\bt{k}_3}_{\bt{k}_1 \bt{k}_2} ({n_{\bt{k}_3}n_{\bt{k}_1}n_{\bt{k}_2}})^{\frac{1}{2}} \left[\frac{Z^{\bt{k}_1}_{\bt{k}_2 \bt{k}_3}}{n_{\bt{k}_1}} + \frac{Z^{\bt{k}_2}_{\bt{k}_3 \bt{k}_1}}{n_{\bt{k}_2}} - \frac{Z^{\bt{k}_3}_{\bt{k}_1  \bt{k}_2}}{n_{\bt{k}_3}}\right]\\
\label{eq:eom_varphi}
&-& {\omega}^{\bt{k}_3}_{\bt{k}_1 \bt{k}_2} + \mathrm{NNTT}^{\bt{k}_3}_{\bt{k}_1 \bt{k}_2}\,,
\vspace{-.2cm}
\end{eqnarray}
where the second equation applies to any triad ($\bt{k}_1 + \bt{k}_2 = \bt{k}_3$). 
 $\mathrm{NNTT}^{\bt{k}_3}_{\bt{k}_1 \bt{k}_2}$ is a short-hand notation for ``nearest-neighbouring-triad terms''; these are nonlinear terms similar to the first line in equation (\ref{eq:eom_varphi}) (see supplemental material). We should emphasise that any dynamical process in the original system results from the dynamics of equations (\ref{eq:eom_n})--(\ref{eq:eom_varphi}).\\

\vspace{-2mm}

\noindent \textbf{Precession resonance.} The nature of the triad phases ${\varphi}^{\bt{k}_3}_{\bt{k}_1 \bt{k}_2}$ is markedly different from the spectrum variables $n_{\bt{k}}$ in that the latter directly contribute to the energy of the system, whereas the former have a contribution that is more subtle. In fact, notice that the RHS of Eq.~(\ref{eq:eom_varphi}) admits, under plausible hypotheses, a zero-mode (in time):
$\Omega^{\bt{k}_3}_{\bt{k}_1 \bt{k}_2} \equiv \lim_{t\to\infty}\frac{1}{t}{\int_0^t{\dot{\varphi}^{\bt{k}_3}_{\bt{k}_1 \bt{k}_2}(t')dt'}}\,.$ This is by definition the \emph{precession frequency} of the triad phase and is a nonlinear function of the variables. Typically it does not perturb the energy dynamics because it is incommensurate with the frequency content of the nonlinear oscillations of the triad variables ${\varphi}^{\bt{k}_3}_{\bt{k}_1 \bt{k}_2}$ and $n_{\bt{k}_1},\,n_{\bt{k}_2},\,n_{\bt{k}_3}.$

However, in special circumstances a resonance occurs whereby the triad precession frequency $\Omega^{\bt{k}_3}_{\bt{k}_1 \bt{k}_2}$ matches one of the typical nonlinear frequencies of the triad variables. In this case, the RHS of Eq.(\ref{eq:eom_n}) will normally develop a zero-mode (in time), leading to a sustained growth of the energy in the corresponding wave spectrum $n_{\bt{k}},$ for some wavevector(s) $\bt{k}.$ We call this a \emph{triad precession/nonlinear frequency} resonance. When several triads are involved in this type of resonance, strong fluxes of enstrophy are exhibited through the network of interconnected triads, leading to coherent collective oscillations and cascades towards small scales.

Now, one can ask: Is this resonance accessible by the manipulation of the initial conditions? The answer is \emph{yes}, and amounts to a simple overall re-scaling of the initial spectrum $n_{\bt{k}} \to \alpha n_{\bt{k}}$ \emph{for all} $\bt{k}$, provided the linear frequency mismatch ${\omega}^{\bt{k}_3}_{\bt{k}_1 \bt{k}_2}$ be nonzero for some triad. To see this, a simple dimensional analysis argument on equation (\ref{eq:eom_n}) shows that under such a re-scaling the nonlinear frequency content re-scales approximately by a factor $\alpha^{\frac{1}{2}}$ (e.g.~enstrophy sets a nonlinear frequency $\sim \sqrt{\mathcal{E}}$). But from equation (\ref{eq:eom_varphi}) the triad precession is the sum of a constant term ($-{\omega}^{\bt{k}_3}_{\bt{k}_1 \bt{k}_2}$) and a term that re-scales by a factor $\alpha^{\frac{1}{2}}.$ Therefore, for some value of $\alpha$ a matching between triad precession and nonlinear frequency can be found, provided ${\omega}^{\bt{k}_3}_{\bt{k}_1 \bt{k}_2} \neq 0.$\\

\vspace{-2mm}

\noindent \textbf{Probing the strong transfer mechanism.} We provide a comprehensive overview of the mechanism by introducing a family of models that interpolates between the original equations (\ref{eq:eom_n})--(\ref{eq:eom_varphi}) and a simple low-dimensional system.
The family is parameterised by two positive numbers $\epsilon_1, \epsilon_2$ which serve to deform the interaction coefficients $Z^{\bt{k}_c}_{\bt{k}_a  \bt{k}_b}.$ The deformation is as follows:

\noindent \textbf{(A)} Choose a triad $\bt{k}_1 +\bt{k}_2 = \bt{k}_3$ with zero frequency mismatch: ${\omega}^{\bt{k}_3}_{\bt{k}_1 \bt{k}_2} = 0.$ This condition is not essential but it leads to simpler analytical solutions for the triad spectrum and triad phase \cite{Kartashova2011}. Do not deform the interaction coefficients for this triad.

\noindent \textbf{(B)} Introduce a fourth mode via the new triad $\bt{k}_2 + \bt{k}_3 = \bt{k}_4.$  This triads' interaction coefficients are replaced by $\epsilon_1 Z^{\bt{k}_c}_{\bt{k}_a  \bt{k}_b}$ so that the limit $\epsilon_1 \to 0$ corresponds to the isolated triad of (A).

\noindent \textbf{(C)} All other triads in the system have their interaction coefficients replaced by $\epsilon_2 Z^{\bt{k}_c}_{\bt{k}_a  \bt{k}_b}.$ The limit $\epsilon_2 \to 0$ recovers the two-triad system in (B).

\noindent \textbf{(D)} The case $\epsilon_1 = \epsilon_2 = 1$ is the full PDE model (\ref{eq:eom_n})--(\ref{eq:eom_varphi}).\\

\vspace{-2mm}

\noindent \textbf{Case (A): $\epsilon_1 = \epsilon_2 = 0.$} This is the isolated triad and is an integrable system \cite{Bustamante2009a}. We choose parameters $F=1, \beta=10,$ and the triad wavevectors $\bt{k}_1 = (1,-4), \,\bt{k}_2 = (1,2), \, \bt{k}_3 = \bt{k}_1 + \bt{k}_2 = (2,-2).$ It follows ${\omega}^{\bt{k}_3}_{\bt{k}_1 \bt{k}_2} = 0.$
The analytical solution for the spectra and triad phase is given in the supplemental material.\\

\vspace{-2mm}

\noindent \textbf{Case (B): $\epsilon_1 \neq 0,\, \epsilon_2 = 0.$} This is a system of two connected triads: $\bt{k}_1 + \bt{k}_ 2 = \bt{k}_3$ and $\bt{k}_2 + \bt{k}_ 3 = \bt{k}_4,$ with $\bt{k}_4 = (3,0).$ The second triad has nonzero frequency mismatch: $\omega^{\bt{k}_4}_{\bt{k}_2  \bt{k}_3} = -8/9.$ The initial conditions for the 6 truly dynamical degrees of freedom are: $\varphi^{\bt{k}_3}_{\bt{k}_1  \bt{k}_2}(0) = \pi/2, \varphi^{\bt{k}_4}_{\bt{k}_2  \bt{k}_3}(0) = -\pi/2$ and $n_{\bt{k}_1}(0) = 5.95984 \times 10^{-5} \alpha, n_{\bt{k}_2}(0) = 1.48858 \times 10^{-3} \alpha, n_{\bt{k}_3} = 1.28792\times 10^{-3} \alpha, n_{\bt{k}_4}(0) = 0,$ where $\alpha$ is a re-scaling parameter. The energy and enstrophy invariants are $E = 0.0215955 \alpha, {\mathcal{E}} = 0.155625 \alpha,$ reducing the effective number of degrees of freedom to $4.$ This is not necessarily integrable, so in general we need to solve the evolution equations (\ref{eq:eom_n})--(\ref{eq:eom_varphi}) numerically in order to study the growth of mode $n_{\bt{k}_4}$ via the precession resonance mechanism. However, in the limit $\epsilon_1 \to 0$ the analytical solution of the isolated (first) triad provides enough information to approximate for the fourth mode's spectrum $n_{\bt{k}_4}$ by quadrature. This leads to an explicit formula for the resonant condition in the form
\begin{equation}
\label{eq:TP-NF resonance}
\Omega^{\bt{k}_4}_{\bt{k}_2  \bt{k}_3} = p \Gamma\,, \quad p \in \mathbb{Z}\,,
\end{equation}
where $\Gamma$ is the isolated triad nonlinear frequency. The isolated triad solution gives $\Gamma = 0.27267 \alpha^{1/2}.$ When $\Omega^{\bt{k}_4}_{\bt{k}_2  \bt{k}_3}$ is such that resonance (\ref{eq:TP-NF resonance}) is nearly satisfied, we have $\Omega^{\bt{k}_4}_{\bt{k}_2  \bt{k}_3} = - 0.20188 \alpha^{1/2} + 8/9$ (see supplemental material), so we obtain
$$\alpha_p = \frac{10.6272}{(0.740382+p)^2}\,, \quad p =0,1, \ldots$$
for predicted values of the initial condition leading to strong growth in $n_{\bt{k}_4}$ in the limit $\epsilon_1 \to 0.$ The case $p<0$ does not apply since it entails shifting the initial phases by $\pi.$

\begin{figure}
\begin{center}
\hspace{-30mm}
\scalebox{0.6}{\input{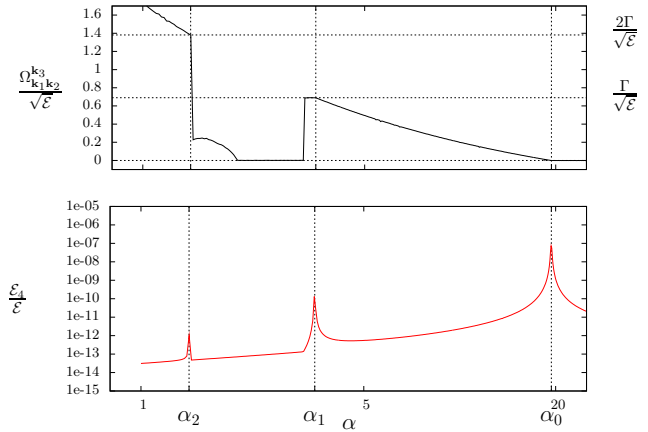}}
\caption{\label{fig:4mode} Numerical results from case (B) with $\epsilon_1=10^{-5}$ showing dimensionless precession (top) and enstrophy transfer efficiency to mode $n_{\bt{k}_4}(t)$ (bottom). Vertical lines indicate predicted resonances (as $\epsilon_1 \to 0$) and show strong transfer efficiency at these values when condition (\ref{eq:TP-NF resonance}) is satisfied (horizontal lines, top figure).}
\end{center}
\end{figure}

\noindent \textbf{Numerical results for case (B): $\epsilon_1 \neq 0, \epsilon_2 = 0.$} We integrate numerically Eqs.~(\ref{eq:eom_n})--(\ref{eq:eom_varphi}) with the above initial conditions, from time $t=0$ to $t=2000/\sqrt{\mathcal{E}}.$ The factor $\sqrt{\mathcal{E}}$ ensures that we compare equivalent nonlinear time scales. Near resonances, strong transfers have a timescale $t \sim 20/\sqrt{\mathcal{E}}$ (including case (C) below); we integrate further in time to obtain well-converged estimates of triad precessions. To confirm the predicted values of $\alpha$ leading to efficient enstrophy transfer towards the fourth mode $n_{\bt{k}_4}(t),$ we define the transfer efficiency as the maximal ratio between this mode's enstrophy and the total, over the simulation time. In the case $\epsilon_1 = 10^{-5}$ we obtain strong peaks of efficiency at the predicted resonant values of $\alpha$ (figure \ref{fig:4mode}, bottom). The precession is best studied in dimensionless form, i.e.~relative to the nonlinear frequency scale $\sqrt{{\mathcal{E}}}$. Figure \ref{fig:4mode} (top) shows the numerically computed dimensionless precession $\Omega^{\bt{k}_4}_{\bt{k}_2  \bt{k}_3}/\sqrt{{\mathcal{E}}}$ as a function of $\alpha,$ leading to a confirmation of the resonances (\ref{eq:TP-NF resonance}) for $p=0,1,2$ (the cases $p=3,4,\ldots$ can be found but have significantly less efficiency). The behaviour of the precession in between resonances is due to a competition between harmonics of the fundamental frequency of the isolated triad equation and can be computed analytically.

When the above resonances are satisfied, trajectories in the 4-dimensional phase space form closed loops corresponding to periodic orbits (see supplemental material). These trajectories visit the neighborhood of a periodic orbit and are ejected along its unstable manifold,
thereby exploring regions of phase space that correspond to high transfer of enstrophy to the fourth mode $n_{\bt{k}_4}.$ These unstable periodic orbits (and other invariant manifolds such as critical points) are \emph{persistent} \cite{Fenichel1971} in parameter space: by varying $\epsilon_1$ from small to large values $\sim 0.1,$ we observe precession resonances and corresponding strong transfers at new values of $\alpha.$ An elementary tracing study of the solution branches in $(\epsilon, \alpha)$ parameter space, using a bisection method (which overlooks possible bifurcations), is presented for the more general case below.\\

\vspace{-2mm}

\begin{figure}
\begin{center}
\hspace{-15mm}
\scalebox{0.7}{\input{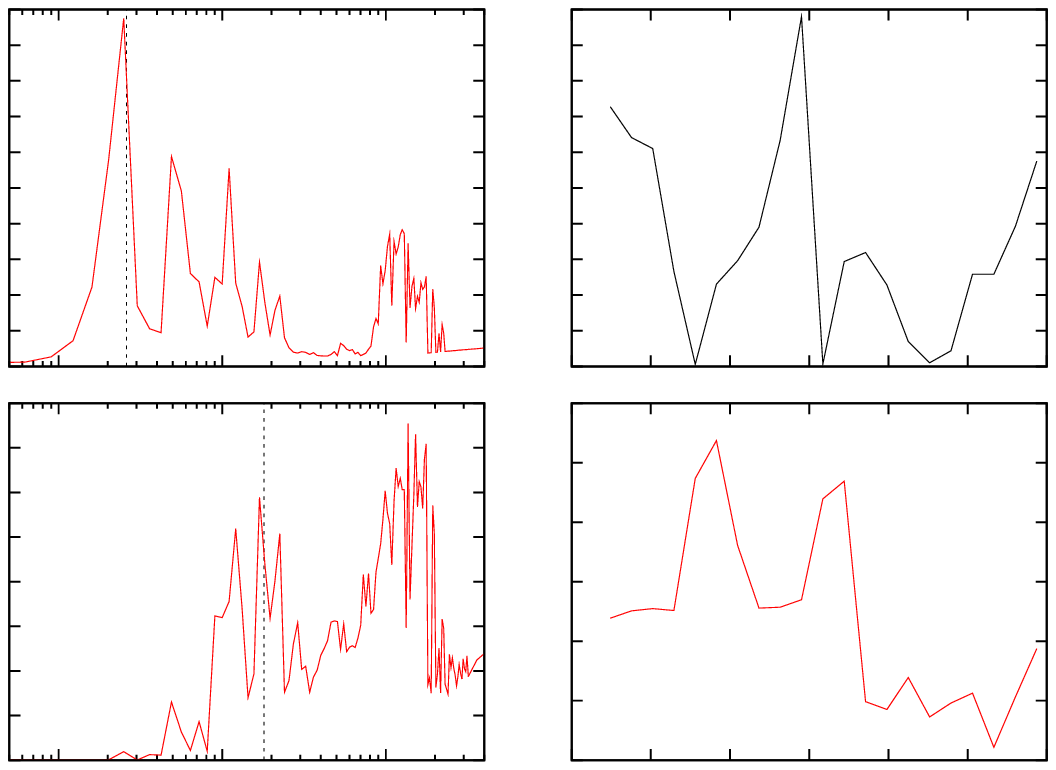}}
\vspace{-3mm}
\caption{\label{fig:eps_bins} Numerical results from case (C) with $\epsilon_1=\epsilon_2=0.1$ at $128^2$ resolution. Left: enstrophy transfer efficiency against $\alpha$ in $bin_2$ and $bin_3.$ Vertical lines denote peaks at $\alpha = 26.01$ and $\alpha = 180.$ Right: dimensionless precession $\Omega^{\bt{c}}_{\bt{a} \bt{b}}/\sqrt{\mathcal{E}}$ for triad $(1,2)+(5,-2)=(6,0)$ and enstrophy transfer efficiency in $bin_3,$ near efficiency peak $\alpha = 180.$}
\end{center}
\end{figure}

\noindent \textbf{Case (C): $\epsilon_1 \neq 0,\, \epsilon_2 \neq 0.$} Next step is to introduce the interactions of additional modes by $\epsilon_2 \neq 0.$ We simulate this via a pseudospectral method with $128^2$ resolution, $2/3$ dealiasing (so $N \approx 2 \times 42^2$) and explicitly introducing the first 4 modes' interactions. We retain the triad initial condition used above and observe the efficiency of enstrophy transfers as $\epsilon_2$ increases from zero. In particular unstable periodic orbits can be traced via bisection in $\alpha.$ We do this tracing until we reach $\epsilon_1 = \epsilon_2 = 0.1,$ which is large enough to give nontrivial transfers to high wavenumbers. The triad initial conditions imply that dynamically populated modes must have wavevectors $\bt{k} = l \bt{k}_1 + m \bt{k_2}\,.$ Thus, we investigate transfer efficiency via a partition of $\bt{k}$-space into bins defined by: $bin_j:$ $2^{j-1} < \sqrt{l^2 + m^2} \leq 2^j\,, \quad j=1,2,3,4.$ Figure \ref{fig:eps_bins} (left) shows the results for the efficiency of enstrophy transfers to $bin_2$ and $bin_3$ as a function of $\alpha.$ The peak at $\alpha \approx 26.01$ for $bin_2$ can be traced back to the predicted resonance $\alpha_0 \approx 19.39$ valid in the limit $\epsilon_1, \epsilon_2 \to 0.$ The remaining peaks correspond to resonances of new modes in the bins. For example, the $bin_3$ peak at $\alpha = 180$ entails a strong enstrophy transfer to mode $(6,0)$ via the precession resonance between triad $(1,2)+(5,-2)=(6,0)$ and the nonlinear oscillations, of the form (\ref{eq:TP-NF resonance}) with $p=0:$ figure \ref{fig:eps_bins} (right) shows a close-up near $\alpha = 180$ of this triad's precession and $bin_3$ efficiency as functions of $\alpha,$ showing that high efficiency corresponds to vanishing precession.

There is significant evidence that the efficiency peaks in figure \ref{fig:eps_bins} correspond to synchronisation of the precession resonances over several triads, as a collective oscillation leading to strong transfers towards small scales. We leave the quantitative study of this synchronisation for the full PDE model case (D), with a general initial condition.\\

\vspace{-2mm}

\begin{figure}
\begin{center}
\hspace{-15mm}
\scalebox{0.7}{\input{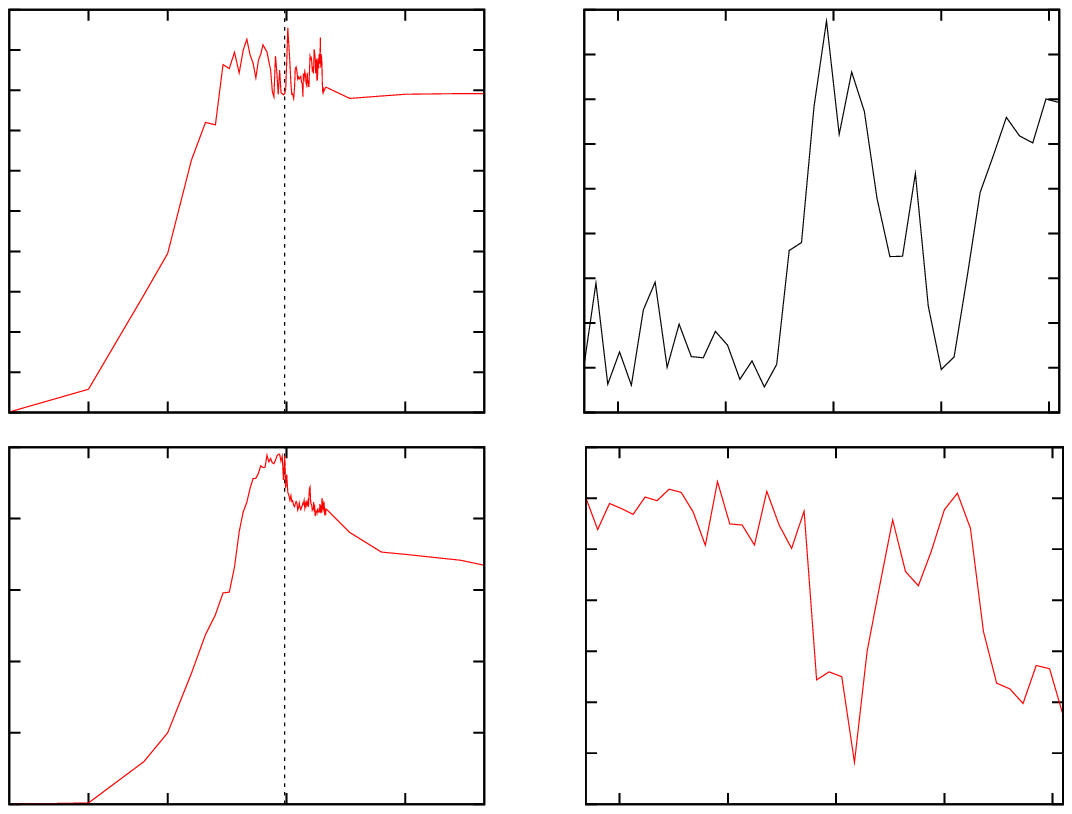}}
\vspace{-3mm}
\caption{\label{fig:full_bins} Numerical results from full PDE model, case (D) at $128^2$ resolution. Left: enstrophy transfer efficiency against $\alpha$ in $Bin_3$ and $Bin_4.$ Vertical lines denote $\alpha = 900.$ Right: dimensionless precession standard deviation (over all interacting triads) and enstrophy transfer efficiency in $Bin_4,$ both near efficiency peak $\alpha = 900.$}
\end{center}
\end{figure}

\noindent \textbf{Case (D): Full PDE (\ref{db_dt}) results, $\epsilon_1 = \epsilon_2 = 1.$} Having shown, in the case of a special initial condition, that strong enstrophy transfers to small scales are due to precession resonances, we consider now a more general large-scale initial condition: $n_{\bt{k}} = 0.032143 \times \alpha |\bt{k}|^{-2}\exp\left(-|\bt{k}|/5\right)$ for $|\bt{k}| \leq 8$ and zero otherwise, where $\alpha$ is the re-scaling parameter. Total enstrophy is ${\mathcal{E}} = 0.155625 \alpha.$ Initial phases $\phi_{\bt{k}}$ are chosen randomly and uniformly between $0$ and $2\pi.$ Direct numerical simulations using the pseudospectral method of case (C) with resolution $128^2$ from $t = 0$ to $t = 800/\sqrt{\mathcal{E}}.$ To study cascades, we partition the $\bt{k}$-space in shell bins defined as follows: $Bin_1:$ $0 < |\bt{k}| \leq 8,$ and $Bin_j:$ $2^{j+1} < |\bt{k}| \leq 2^{j+2} \quad j=2,3, \ldots,$ so nonlinear interactions lead to successive transfers $Bin_1 \to Bin_2 \to Bin_3 \to Bin_4.$ Near resonances, strong transfers to $Bin_4$ have a timescale $t \sim 40/\sqrt{\mathcal{E}}.$ Figure \ref{fig:full_bins} (left) shows the efficiencies of enstrophy transfers from $Bin_1$ to $Bin_3$ and $Bin_4.$ Peaks concentrate in a broad region, corresponding to collectively synchronised precession resonances. Strong synchronisation is signalled by minima of the dimensionless precession standard deviation $\sigma = \sqrt{\langle \Omega^2\rangle - \langle \Omega\rangle^2}/\sqrt{\mathcal{E}}$ averaged over the whole set of triad precessions (about $N^2$ triads, not just the $N-2$ independent precessions). A close-up around the $Bin_4$ efficiency peak at $\alpha \sim 900$ is shown in figure \ref{fig:full_bins} (right), showing that efficiency peaks correspond to minima of $\sigma.$

\begin{figure}
\begin{center}
\hspace{-7mm}
\scalebox{0.66}{\input{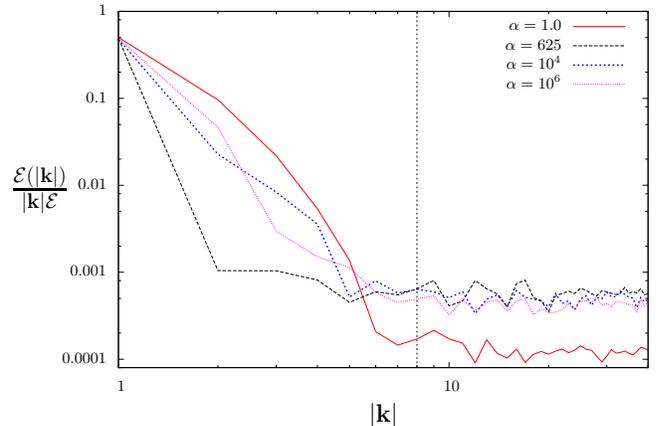}}
\vspace{-8mm}
\caption{\label{fig:Aspec} Time-averaged dimensionless enstrophy spectra, compensated for enstrophy equipartition, for various values of $\alpha$ from case (D) at $128^2$ resolution.}
\end{center}
\end{figure}

\noindent \textbf{Enstrophy fluxes, equipartition and resolution study.} Figure \ref{fig:Aspec} shows, for representative values of $\alpha,$ time averages ($T = 800/\sqrt{\mathcal{E}}$) of dimensionless enstrophy spectra ${\mathcal{E}}_{k}/{\mathcal{E}},$ compensated for enstrophy equipartition to aid visualisation. In all cases the system reaches small-scale equipartition ($Bin_2$--$Bin_4$) quite soon: $T_{\mathrm{eq}}\approx 80/\sqrt{\mathcal{E}}$. Remarkably, the flux of enstrophy from large scales ($Bin_1$) to small scales ($Bin_4$) is $50\%$ greater in the resonant case ($\alpha = 625$) than in the limit of very large amplitudes ($\alpha = 10^6$). Also, in the resonant case equipartition encroaches on the large scales. At double the resolution ($256^2$), the enstrophy cascade goes further to $Bin_5$ and all above analyses are verified, with $Bin_4$ replaced by $Bin_5.$ The transfer time scales seem to increase weakly with resolution, in accordance with known results \cite{tran2006,lopes2006,dritschel2007}.

\noindent \textbf{Conclusions and Extensions.}
There is vast literature on precession-like resonances in galactic dynamics, notably Pluto precession-orbit resonance and orbital 2:5 Saturn-Jupiter resonance \cite{rubincam2000pluto,michtchenko2001modeling}. Critical balance turbulence principle \cite{Goldreich1995,Nazarenko2011a} is effectively satisfied at the resonance (\ref{eq:TP-NF resonance}), where we fine-tune a nonlinear frequency (the nonlinear contributions to $\Omega^{\bt{k}_3}_{\bt{k}_1 \bt{k}_2}$, see equation (\ref{eq:eom_varphi})) with the linear frequency mismatch $\omega^{\bt{k}_3}_{\bt{k}_1 \bt{k}_2}$. Possibility for future work includes investigating this precession resonance mechanism in more complex triad systems (water gravity waves, magneto-hydrodynamics) and quartet and higher-order systems (Kelvin waves in superfluids, nonlinear optics), along with including forcing and dissipation.

\section{ACKNOWLEDGMENTS}
We thank C. Connaughton, F. Dias, J. Dudley, P. Lynch, B. Murray, S. Nazarenko and A. Newell for useful discussions.
We are deeply grateful to the organisers of the ``Thematic Program on the
Mathematics of Oceans'' at the Fields Institute, Toronto, who provided support
and use of facilities for this research. Additional support for this work was
provided by UCD Seed Funding project SF652,  IRC Fellowship ``The nonlinear
evolution of phases and energy cascades in discrete wave turbulence'', Science Foundation Ireland (SFI) under Grant Number 12/IP/1491, and ERC-2011-AdG 290562-MULTIWAVE.

\bibliography{bibliography}
\end{document}